# Dual To Ratio Cum Product Estimator In Stratified Random Sampling

Rajesh Singh, Mukesh Kumar and Manoj K. Chaudhary

Department of Statistics, B.H.U., Varanasi (U.P.),India

Cem Kadilar

Department of Statistics, Hacettepe University,

Beytepe 06800, Ankara, Turkey

**Abstract**

Tracy et al.[8] have introduced a family of estimators using Srivenkataramana and Tracy ([6],[7]) transformation in simple random sampling. In this article, we have proposed a dual to ratio-cum-product estimator in stratified random sampling. The expressions of the mean square error of the proposed estimators are derived. Also, the theoretical findings are supported by a numerical example.

## 1. Introduction

In planning surveys, stratified sampling has often proved as useful in improving the precision of un-stratified sampling strategies to estimate the finite population mean of the study variable, $\overline{Y} = \frac{1}{N} \sum_{h=1}^{L} \sum_{i=1}^{N_h} y_{hi}$. Let y, x and z respectively, be the study and auxiliary variates on each unit $U_h$ (h=1,2,3, ---, N) of the population U. Here the size of the stratum $U_h$ is $N_h$, and the size of simple random sample in stratum $U_h$ is $n_h$, where h=1, 2,---,L. In this study, under stratified random sampling without replacement scheme, we suggest estimators to estimate $\overline{Y}$ by considering the estimators in Plikusas [3] and in Tracy et al. [8].

To obtain the bias and MSE of the proposed estimators, we use the following notations in the rest of the article:

$$\bar{y}_{st} = \sum_{h=1}^{L} w_h \bar{y}_h = \bar{Y}(1+e_0),$$

$$\bar{x}_{st} = \sum_{h=1}^{L} w_h \bar{x}_h = \bar{X}(1+e_1),$$

$$\bar{z}_{st} = \sum_{h=1}^{L} w_h \bar{z}_h = \bar{Z}(1+e_2),$$

where, $w_h = \dfrac{N_h}{N}$.

Such that,

$$E(e_0) = E(e_1) = E(e_2) = 0,$$

$$V_{rst} = \sum_{h=1}^{L} w_h^{r+s+t} \frac{E[(\bar{y}_h - \bar{Y}_h)^r (\bar{x}_h - \bar{X}_h)^s (\bar{z}_h - \bar{Z}_h)^t]}{\bar{Y}^r \bar{X}^s \bar{Z}^t}. \quad (1)$$

where $\bar{y}_h$ and $\bar{Y}_h$ are the sample and population means of the study variable in the stratum h, respectively. Similar expressions for X and Z can also be defined.

Using (1), we can write

$$E(e_0^2) = \frac{\sum_{h=1}^{L} w_h^2 \gamma_h S_{yh}^2}{\bar{Y}^2} = V_{200},$$

$$E(e_1^2) = \frac{\sum_{h=1}^{L} w_h^2 \gamma_h S_{xh}^2}{\bar{X}^2} = V_{020},$$

$$E(e_2^2) = \frac{\sum_{h=1}^{L} w_h^2 \gamma_h S_{zh}^2}{\bar{Z}^2} = V_{002},$$

$$E(e_0 e_1) = \frac{\sum_{h=1}^{L} w_h^2 \gamma_h S_{xyh}}{\bar{X}\bar{Y}} = V_{110},$$

$$E(e_1 e_2) = \frac{\sum_{h=1}^{L} w_h^2 \gamma_h S_{xzh}}{\bar{X}\bar{Z}} = V_{011},$$

where

$$S_{yh}^2 = \frac{\sum_{i=1}^{N_h}(\bar{y}_h - \bar{Y}_h)^2}{N_h - 1}, \quad S_{xh}^2 = \frac{\sum_{i=1}^{N_h}(\bar{x}_h - \bar{X}_h)^2}{N_h - 1}, \quad S_{zh}^2 = \frac{\sum_{i=1}^{N_h}(\bar{z}_h - \bar{Z}_h)^2}{N_h - 1},$$

$$S_{xyh} = \frac{\sum_{i=1}^{N_h}(\bar{x}_h - \bar{X}_h)(\bar{y}_h - \bar{Y}_h)}{N_h - 1}, \quad S_{yzh} = \frac{\sum_{i=1}^{N_h}(\bar{y}_h - \bar{Y}_h)(\bar{z}_h - \bar{Z}_h)}{N_h - 1},$$

$$S_{xzh} = \frac{\sum_{i=1}^{N_h}(\bar{x}_h - \bar{X}_h)(\bar{z}_h - \bar{Z}_h)}{N_h - 1}, \quad \gamma_h = \frac{1 - f_h}{n_h}, \quad f_h = \frac{n_h}{N_h}, \quad w_h = \frac{N_h}{N}.$$

The combined ratio and the combined product estimators are, respectively, defined as

$$\bar{y}_1 = \bar{y}_{st}\left(\frac{\bar{X}}{\bar{x}_{st}}\right), \tag{2}$$

$$\bar{y}_2 = \bar{y}_{st}\left(\frac{\bar{z}_{st}}{\bar{Z}}\right) \tag{3}$$

And the MSE of $\bar{y}_1$ and $\bar{y}_2$ to the first degree of approximation are, respectively, given by

#

$$MSE(\bar{y}_1) \cong \bar{Y}^2(V_{200} + V_{020} - 2V_{110}) \tag{4}$$

$$MSE(\bar{y}_2) \cong \bar{Y}^2(V_{200} + V_{020} + 2V_{110}) \tag{5}$$

Note that $\bar{Y} = \bar{Y}_{st} = \sum_{h=1}^{L} w_h \bar{Y}_h$. Similar expressions for X and Z can also be defined.

## 2. Classical Estimators

Srivenkataramana and Tracy ([6],[7]) considered a simple transformation as

$$u_i = A - x_i, \quad (i=1, 2, \ldots, N)$$

$$\rightarrow \bar{u} = A - \bar{x},$$

where A is a scalar to be chosen. This transformation renders the situation suitable for a product method instead of ratio method. Clearly $\bar{u}_{st}(= A - \bar{x}_{st})$ is unbiased for $U(= A - \bar{X})$. Using this transformation, an estimator in the stratified random sampling is defined as

$$\bar{y}_3 = \bar{y}_{st}\left(\frac{\bar{u}_{st}}{U}\right) \qquad (6)$$

This is a product type estimator (alternative to combined ratio type estimator) in stratified random sampling.

The exact expression for MSE of $\bar{y}_3$ is given by

$$MSE(\bar{y}_3) = \bar{Y}^2(V_{200} + \theta^2 V_{020} - 2\theta V_{110}) \qquad (7)$$

where $\theta = \dfrac{\bar{X}}{(A - \bar{X})}$.

In some survey situations, information on a second auxiliary variable, Z, correlated negatively with the study variable, Y, is readily available. Let $\bar{Z}$ be the known population mean of Z. To estimate $\bar{Y}$, Singh[4] considered ratio-cum-product estimator as

$$\bar{y}_4 = \bar{y}\left(\frac{\bar{X}}{\bar{x}}\right)\left(\frac{\bar{z}}{\bar{Z}}\right),$$

where Perri[2] used $\bar{t}_X = \bar{x} + \alpha(\bar{X} - \bar{x})$ and $\bar{t}_Z = \bar{z} + \beta(\bar{Z} - \bar{z})$ instead of $\bar{x}$ and $\bar{z}$, respectively. Here, $\alpha$ and $\beta$ are constants that make the MSE minimum.

Adapting $\bar{y}_4$ to the stratified random sampling, the ratio cum product estimator using two auxiliary variables can be defined as

$$\bar{y}_5 = \bar{y}_{st}\left(\frac{\bar{X}}{\bar{x}_{st}}\right)\left(\frac{\bar{z}_{st}}{\bar{Z}}\right) \qquad (8)$$

The approximate MSE of this estimator is

$$\text{MSE}(\bar{y}_6) \cong \bar{Y}^2[V_{200} + V_{020} + V_{002} + 2(V_{101} - V_{110} - V_{011})] \tag{9}$$

## 3. Suggested Estimators

Tracy et al. [8] introduced a product estimator using two auxiliary variables in the simple random sampling given by

$$\bar{y}_6 = \bar{y}\left(\frac{\bar{u}}{\bar{U}}\right)\left(\frac{\bar{z}}{\bar{Z}}\right) \tag{10}$$

Motivated by Tracy et al. [8], we propose the following product estimator for the stratified random sampling scheme as

$$\bar{y}_7 = \bar{y}_{st}\left(\frac{\bar{u}_{st}}{\bar{U}}\right)\left(\frac{\bar{z}_{st}}{\bar{Z}}\right) \tag{11}$$

Expressing $\bar{y}_7$ in terms of e's, we can write (11) as

$$\bar{y}_7 = \bar{Y}(1+e_0)(1-\theta e_1)(1+e_2)$$

The MSE$(\bar{y}_7)$ to the first order of approximation, is given as

$$\text{MSE}(\bar{y}_7) = \bar{Y}^2[V_{200} + \theta^2 V_{020} + V_{002} - 2(\theta V_{110} - V_{101} + \theta V_{011})] \tag{12}$$

and this MSE equation is minimised for

$$\theta = \frac{V_{110} + V_{011}}{V_{020}} = \theta_{opt}(\text{say})$$

Note that the corresponding A is

$$A_{opt} = \frac{(1-\theta_{opt})\bar{X}}{\theta_{opt}}.$$

By putting the optimum value of $\theta$ in (12), we can obtain the minimum MSE equation for the first proposed estimator, $\bar{y}_7$.

**Remark 3.1** : The value of $\bar{X}$ is known, but the exact values of $V_{110}$, $V_{011}$ and $V_{020}$ are rarely available in practice. However in repeated surveys or studies based on multiphase sampling, where information is gathered on several occasions it may be possible to guess the values of $V_{110}$, $V_{011}$ and $V_{020}$ quite accurately. Even though this approach may reduce the precision, it may be satisfactory provided the relative decrease in precision is marginal, see Tracy et al. [8].

Plikusas[3] defined dual to ratio cum product estimator in stratified random sampling as

$$\bar{y}_8 = \bar{y}_{st} \frac{\bar{x}_{st}^*}{\bar{X}} \cdot \frac{\bar{Z}}{\bar{z}_{st}^*} \qquad (13)$$

where

$$\bar{x}_{st}^* = (1 + g_h \bar{X}_h) - g_h \bar{x}_h, \quad \bar{z}_{st}^* = (1 + g_h \bar{Z}_h) - g_h \bar{z}_h.$$

and $g_h = \dfrac{n_h}{(N_h - n_h)}$.

Considering the estimator in (13) and motivated by Singh et al. [5], we propose a family of dual to ratio cum product estimator as –

$$\bar{y}_9 = \bar{y}_{st} \left(\frac{\bar{x}_{st}^*}{\bar{X}}\right)^{\alpha_4} \left(\frac{\bar{Z}}{\bar{z}_{st}^*}\right)^{\alpha_5} \qquad (14)$$

To obtain the MSE of the second proposed estimator, $\bar{y}_9$, we write

$$\bar{y}_{st} = \bar{Y}(1 + e_0),$$

$$\bar{x}_{st}^* = (1 + e_1'),$$

$$\bar{z}_{st}^* = (1 + e_2').$$

Expressing (14) in terms of e's, we have

$$\bar{y}_9 = \bar{Y}(1 + e_0)(1 + e_1')^{\alpha_1}(1 + e_2')^{-\alpha_2} \qquad (15)$$

Expanding the right hand side of (15), to the first order of approximation, we get

$$(\bar{y}_9 - \bar{Y}) = \bar{Y}[e_0 + \alpha_1 e_0 e_1' - \alpha_2 e_0 e_2' + \alpha_1 e_1' - \alpha_2 e_2' - \alpha_1 \alpha_2 e_1' e_2'$$

$$+ \frac{\alpha_1(\alpha_1 - 1)}{2}e_1'^2 + \frac{\alpha_2(\alpha_2 + 1)}{2}e_2'^2] \qquad (16)$$

Squaring both sides of (16) and then taking expectation, we obtain the MSE of the second proposed estimator, $\bar{y}_9$, to the first order approximation, as

$$MSE(\bar{y}_9) \cong \bar{Y}^2\{V_{200}' + \alpha_1^2 V_{020}' + \alpha_2^2 V_{002}' + 2(\alpha_1 V_{110}' - \alpha_1 \alpha_2 V_{011}' - \alpha_2 V_{101}')\} \qquad (17)$$

where

$$V_{rst}' = \sum_{h=1}^{L} \frac{w^{r+s+t}(-g)^{s+t} E(\bar{y}_h - \bar{Y}_h)(\bar{x}_h - \bar{X}_h)(\bar{z}_h - \bar{Z}_h)}{\bar{Y}^r \bar{X}^s \bar{Z}^t} \qquad (18)$$

This MSE equation is minimized for the optimum values of $\alpha_1$ and $\alpha_2$ given by

$$\alpha_1' = \frac{V_{101}' V_{011}' - V_{110}' V_{002}'}{V_{020}' V_{002}' - V_{011}'^2} \qquad (19)$$

$$\alpha_2' = \frac{V_{020}' V_{101}' - V_{110}' V_{011}'}{V_{020}' V_{002}' - V_{011}'^2} \qquad (20)$$

Putting these values of $\alpha_1'$ and $\alpha_2'$ in MSE ($\bar{y}_9$), given in (17), we obtain the minimum MSE of the second proposed estimator, $\bar{y}_9$.

## 4. Theoretical Efficiency Comparisons

In this section, we first compare the efficiency between the first proposed estimator, $\bar{y}_7$, with the classical combined estimator, $\bar{y}_{st}$, as follows:

$$MSE(\bar{y}_7) < V(\bar{y}_{st})$$

$$Y^2[V_{200} + \theta^2 V_{020} + V_{002} - 2(\theta V_{110} - V_{101} + \theta V_{011})] < Y^2 V_{200}.$$

The estimator $\bar{y}_7$ is better than the usual estimator $\bar{y}_{st}$, if and only if,

$$\frac{B_1}{2B_2} < 1, \qquad (21)$$

where, $B_1 = \theta^2 V_{020} + V_{002}$ and $B_2 = \theta V_{110} - V_{101} + \theta V_{011}$.

If the condition (21) is satisfied, the first proposed estimator, $\bar{y}_7$, performs better than the classical combined estimator.

We also find the condition under which the second proposed estimator, $\bar{y}_9$, performs better than the classical combined estimator in theory as follows:

$$MSE(\bar{y}_9) < V(\bar{y}_{st}),$$

$$Y^2\{V'_{200} + \alpha_1^2 V'_{020} + \alpha_2^2 V'_{002} + 2(\alpha_1 V'_{110} - \alpha_1 \alpha_2 V'_{011} - \alpha_2 V'_{101})\} < Y^2 V_{200},$$

$$\frac{\alpha_1^2 V'_{020} + \alpha_2^2 V'_{002} - 2\alpha_1 \alpha_2 V'_{011}}{\alpha_2 V'_{101} - \alpha_1 V'_{110}} < 1,$$

The estimator $\bar{y}_9$ is better than the usual estimator $\bar{y}_{st}$, if and only if,

$$\frac{C}{D} < 1, \qquad (22)$$

where, $C = \alpha_1^2 V'_{020} + \alpha_2^2 V'_{002} - 2\alpha_1 \alpha_2 V'_{011}$ and $D = \alpha_2 V'_{101} - \alpha_1 V'_{110}$.

## 5. Numerical Example

In this section, we use the data set earlier used in Koyuncu and Kadilar[1]. The population statistics are given in Table 1. In this data set, the study variable (Y) is the number of teachers, the first auxiliary variable (X) is the number of students, and the second auxiliary variable (Z) is the number of classes in both primary and secondary schools for 923 districts at 6 regions ( as 1: Marmara, 2: Agean, 3: Mediterranean, 4: Central Anatolia, 5: Black Sea, 6: East and South east Anatolia) in Turkey in 2007, see Koyuncu and Kadilar[1]. Koyuncu and Kadilar[1] have used Neyman allocation for allocating the samples to different strata. Note that all correlations between the study and auxiliary variables are positive. Therefore, we decide not to use product estimators for this data set for efficiency comparison. For this reason, we apply the classical combined estimator, $\bar{y}_{st}$, combined ratio estimator, $\bar{y}_1$, the ratio-cum-product estimator, $\bar{y}_5$, Plikusas [3] estimator, $\bar{y}_8$, and the second proposed estimator, $\bar{y}_9$, to the data set. For the efficiency comparison, we compute percent relative efficiencies as

$$\text{PRE}(\bar{y}_i) = \frac{\text{MSE}(\bar{y}_{st})}{\text{MSE}(\bar{y}_i)} \times 100, \quad i = st, 1, 5, 8, 9.$$

**Table 1**. Data Statistics of Population

| | | |
|---|---|---|
| $N_1=127$ | $N_2=117$ | $N_3=103$ |
| $N_4=170$ | $N_5=205$ | $N_6=201$ |
| $n_1=31$ | $n_2=21$ | $n_3=29$ |
| $n_4=38$ | $n_5=22$ | $n_6=39$ |
| $S_{y1}=883.835$ | $S_{y2}=644$ | $S_{y3}=1033.467$ |
| $S_{y4}=810.585$ | $S_{y5}=403.654$ | $S_{y6}=711.723$ |
| $\bar{Y}_1=703.74$ | $\bar{Y}_2=413$ | $\bar{Y}_3=573.17$ |
| $\bar{Y}_4=424.66$ | $\bar{Y}_5=267.03$ | $\bar{Y}_6=393.84$ |
| $S_{x1}=30486.751$ | $S_{x2}=15180.760$ | $S_{x3}=27549.697$ |
| $S_{x4}=18218.931$ | $S_{x5}=8997.776$ | $S_{x6}=23094.141$ |
| $\bar{X}_1=20804.59$ | $\bar{X}_2=9211.79$ | $\bar{X}_3=14309.30$ |
| $\bar{X}_4=9478.85$ | $\bar{X}_5=5569.95$ | $\bar{X}_6=12997.59$ |
| $S_{xy1}=25237153.52$ | $S_{xy2}=9747942.85$ | $S_{xy3}=28294397.04$ |
| $S_{xy1}=14523885.53$ | $S_{xy1}=3393591.75$ | $S_{xy6}=15864573.97$ |
| $\rho_{xy1}=0.936$ | $\rho_{xy2}=0.996$ | $\rho_{xy3}=0.994$ |
| $\rho_{xy4}=0.983$ | $\rho_{xy5}=0.989$ | $\rho_{xy6}=0.965$ |
| $S_{z1}=555.5816$ | $S_{z2}=365.4576$ | $S_{z3}=612.9509281$ |
| $S_{z4}=458.0282$ | $S_{z5}=260.8511$ | $S_{z6}=397.0481$ |
| $\bar{Z}_1=498.28$ | $\bar{Z}_2=318.33$ | $\bar{Z}_3=431.36$ |
| $\bar{Z}_4=498.28$ | $\bar{Z}_5=227.20$ | $\bar{Z}_6=313.71$ |
| $S_{yz1}=480688.2$ | $S_{yz2}=230092.8$ | $S_{yz1}=623019.3$ |
| $S_{yz1}=364943.4$ | $S_{yz1}=101539$ | $S_{yz1}=277696.1$ |
| $S_{xz1}=15914648$ | $S_{xz2}=5379190$ | $S_{xz3}=164900674.56$ |
| $S_{xz4}=8041254$ | $S_{xz5}=2144057$ | $S_{xz1}=8857729$ |
| $\rho_{yz1}=0.978914$ | $\rho_{yz2}=0.9762$ | $\rho_{y3}=0.983511$ |

$\rho_{yz4} = 0.982958$   $\rho_{yz5} = 0.964342$   $\rho_{yz1} = 0.982689$

**Table 2.** Percent Relative Efficiencies (PRE) of estimators

| Estimators | Values of $\alpha_1$ | Values of $\alpha_2$ | PRE($\bar{y}_i$) |
|---|---|---|---|
| $\bar{y}_{st}$ | 0 | 0 | 100 |
| $\bar{y}_1$ | 1 | 0 | 1029.469 |
| $\bar{y}_5$ | 1 | 1 | 149.686 |
| $\bar{y}_8$ | 1 | 1 | 115.189 |
| MSE($\bar{y}_9$)$_{min}$ | 6.2918 | -0.8870 | 2854.549 |

**Table 3.** The MSE values according to A

| Value of $\theta$ | Corresponding value of A | MSE($\bar{y}_7$) |
|---|---|---|
| <0.8 | - | >V(yst) |
| 0.8 | 25779.79 | 2186.879 |
| 0.9 | 24188.44 | 1814.999 |
| 1.00 | 22915.37 | 1492.895 |
| 1.10 | 21873.76 | 1220.564 |
| 1.20 | 21005.75 | 998.009 |
| 1.30 | 20271.29 | 825.227 |
| 1.40 | 19641.74 | 702.221 |
| 1.50 | 19096.14 | 628.989 |
| *1.5971(opt)* | *18631.62(opt)* | *605.511\** |
| 1.60 | 18618.74 | 605.532 |
| 1.70 | 18197.50 | 631.849 |
| 1.80 | 17823.06 | 707.941 |
| 1.90 | 17488.04 | 833.807 |
| 2.00 | 17186.53 | 1009.448 |
| 2.10 | 16913.72 | 1234.864 |
| 2.20 | 16665.72 | 1510.054 |
| 2.30 | 16439.29 | 1835.019 |
| 2.40 | 16231.72 | 2209.758 |
| >2.40 | - | >V(yst) |

\* MSE (min) at the value A(optimal).

## 6. Conclusion

When we examine Table 2, we observe that the second proposed estimator, $\bar{y}_9$, under optimum condition certainly performs quite better than all other estimators discussed here. Although the correlations are negative, we also examine the performance of the first proposed estimator, $\bar{y}_7$, according to the classical combined estimator. Therefore, for various values of A and $\theta$ in Table 3, the MSE values of $\bar{y}_{st}$ and $\bar{y}_7$ are computed. From Table 3, we observe that the first proposed estimator, $\bar{y}_7$, performs better than the estimator, $\bar{y}_{st}$, for a wide range of $\theta$ as $\theta \in [0.8, 2.40]$, even in the negative correlations.

## Acknowledgements

The second author (Mukesh Kumar) is thankful to UGC, New Delhi, India, for providing financial assistance. The authors would like to thank the referee for his constructive suggestions on an earlier draft of the paper.

## References

[1] Koyuncu, N. and Kadilar, C. Family of Estimators of Population Mean Using Two Auxiliary Variables in Stratified Random Sampling Commun. in Statist.—Theor. and Meth, 38, 2009, 2398–2417.

[2] Perri, P.F. Improved ratio-cum-product type estimators. Statist. In Trans, 2007, 851-69.

[3] Plikusas, A. Some overview of the ratio type estimators In: Workshop on survey sampling theory and methodology, Statistics Estonia, 2008.

[4] Singh, M. P. Ratio-cum-product method of estimation. Metrika 12, 1967, 34 -42.

[5] Singh, R., Kumar, M., Chauhan, P., Sawan, N. and Smarandache, F. A general family of dual to ratio-cum-product estimator in sample surveys. Statist. In Trans- New series. IJSA, 2012, 1(1), 101-109.

[6] Srivenkataramana, T. and Tracy, D.S. An alternative to ratio method in sample surveys. Ann. Inst. Statist. Math.32 A, 1980, 111-120.

[7] Srivenkataramana, T. and Tracy, D.S. Extending product method of estimation to positive correlation case in surveys. Austral. J. Statist. 23, 1981, 95-100.

[8] Tracy, D.S., Singh, H.P. and Singh, R. An alternative to the ratio-cum-product estimator in sample surveys. Jour. of Statist. Plann. and Infere. 53, 1996, 375 -387.